\documentclass[aps,amsmath,amssymb,pre,reprint]{revtex4-1}
\usepackage{hyperref}
\usepackage{graphicx}
\usepackage{pdfpages}

\newcommand{\sqb}[1]{\left[ #1 \right]}
\newcommand{\rndb}[1]{\left( #1 \right)}

\newcommand{\norm}[1]{\left| #1 \right|}
\newcommand{\vect}[1]{\boldsymbol{#1}}
\newcommand{\tens}[1]{\underline{\underline{#1}}}

\newcommand{\divv}{\nabla \cdot}

\newcommand{\act}{\zeta}
\newcommand{\actc}{\zeta_c}
\newcommand{\actt}{\tilde{\zeta}}

\begin{document}

\title{Instabilities, motion and deformation of active fluid droplets}
\author{Carl A. Whitfield}
\email[Corresponding author: ]{carl.whitfield@physics.org}
\altaffiliation[Current address: ]{Department of Physics, University of Warwick, Coventry, UK, CV3 7AL}
\affiliation{Department of Physics and Astronomy, University of Sheffield, Sheffield, UK, S3 7RH}
\author{Rhoda J. Hawkins}
\affiliation{Department of Physics and Astronomy, University of Sheffield, Sheffield, UK, S3 7RH}
\date{\today}
\begin{abstract}
We consider two minimal models of active fluid droplets that exhibit complex dynamics including steady motion, deformation, rotation and oscillating motion. First we consider a droplet with a concentration of active contractile matter adsorbed to its boundary. We analytically predict activity driven instabilities in the concentration profile, and compare them to the dynamics we find from simulations. Secondly, we consider a droplet of active polar fluid of constant concentration. In this system we predict, motion and deformation of the droplets in certain activity ranges due to instabilities in the polarisation field. Both these systems show spontaneous transitions to motility and deformation which resemble dynamics of the cell cytoskeleton in animal cells.
\end{abstract}
\maketitle

\section{Introduction}
\label{sec:intro}

In animal cells, motility and morphology are strongly coupled and are largely due to the activity of the cell cytoskeleton. Research into these areas is broad and has many applications, from studying metastatic cancer cells to wound healing. In order to mimic aspects of these systems we model, both analytically and numerically, examples of active cytoskeletal material confined to droplets. An active material is defined as driven out-of-equilibrium by the internal energy of its constituent particles \cite{Marchetti2013}. We use the hydrodynamic model of an active polar fluid outlined in \cite{Kruse2004,Kruse2005,Furthauer2012} to model the behaviour of such a material at long length and time scales.

Over the past decade there have been a number of calculations of instabilities and non-equilibrium steady states in active liquid crystals; thin or 2D flat films \cite{Kruse2004,Voituriez2005,Voituriez2006,Kruse2006,Bois2011,Sarkar2015a}, thin cortical layers \cite{Zumdieck2005,Hawkins2011,Joanny2013,Khoromskaia2015}, confined in emulsion droplets or vesicles \cite{Callan-Jones2008a,Tjhung2012,Joanny2012a,Blanch-Mercader2013,Giomi2014,Whitfield2014,Marth2014,Tjhung2015}, and simplified models of animal and plant cells \cite{Hawkins2009,Woodhouse2012,Callan-Jones2013,Kumar2014,Turlier2014,Callan-Jones2016}. {In this paper we model deforming active droplets immersed in a passive fluid using linear perturbation theory. By making justified assumptions, we are able to predict non-equilibrium phase transitions in both of the systems we consider, and predict how the droplet deformation couples to these. These analytical calculations are presented for the three-dimensional case and also repeated for the two-dimensional analogue where we find qualitatively similar results. Numerical simulations use the two-dimensional Immersed Boundary method used in \cite{Whitfield2016} and are directly compared to the two-dimensional analytical calculation.

The models presented here are relevant to active systems \emph{in vitro} (constructed using techniques in \cite{Bendix2008, Sanchez2012, Keber2014}) as well mimicking aspects of cell dynamics. The two cases we consider correspond to two limits of active cytoskeletal behaviour (see figure \ref{fig:sketch}) that represent the minimum degrees of freedom required to observe interesting out-of-equilibrium dynamics. In both cases we consider a 1-component model used originally in \cite{Kruse2004}, which allows us to investigate the coupling with droplet shape dynamics analytically. The linear stability analyses are restricted by assumptions which enable an analytical understanding of the mechanisms involved in producing the observed behaviour in numerical simulations.
\begin{figure}[h!]
	\hfill \includegraphics[width=\columnwidth]{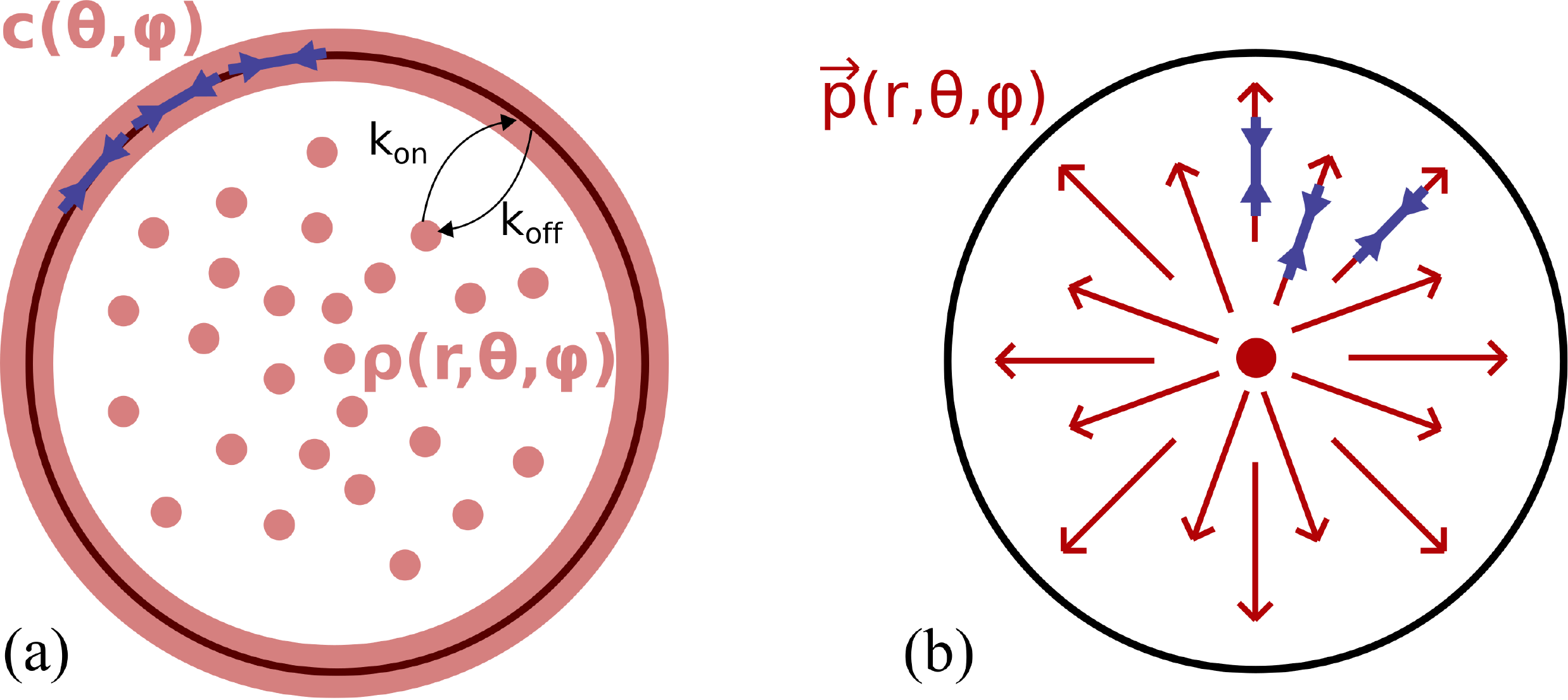}
	\caption{2D schematic of \textbf{(a)} Active fluid interface: active concentration $c$ on the droplet interface coupled to the internal concentration $\rho$. \textbf{(b)} Active polar droplet: constant density of active filaments with local average polarisation $\vect{p}$ (red arrows). Blue arrows indicate active contractile force dipoles.}
	\label{fig:sketch}
\end{figure}

Firstly, we consider an isotropic layer of contractile active material confined to an interface between two fluids, which has physical similarities to the actomyosin cortex in cells. The stresses generated are confined to the plane of the interface giving rise to flows in the surrounding fluid and deformation of the interface itself. Interestingly, diffusion of the active particles through the bulk can result in { a change in which mode of the perturbation has lowest critical activity, from a single peak instability driving droplet motion to higher modes which produce symmetric deformation}. Furthermore, simulations show that advection through the bulk can stabilise such modes. This suggests that droplets with an active interface could spontaneously deform and possibly divide due to the feedback from the fluid flow. 

Secondly, we consider a highly ordered active polar liquid crystal confined inside a fluid droplet. In this case the polarisation gradients direct the internal stresses giving rise to fluid flow. A polar anchoring condition at the interface means that the deformation of the droplet and polarisation field are strongly coupled. We find in this case there is a separation of swimming and stationary deforming modes, such that extensile activity destabilises the defect position and results in a swimming drop, whereas a contractile activity stabilises the centred defect position and gives rise to deformations of the interface.

\section{Active Fluid Interface}
\label{sec:model}

In this section we consider a fluid droplet coated by active particles on its interface that are isotropically ordered. Such systems have been found to self-organise in \emph{in vitro} experiments using reconstituted active cytoskeletal material contained in vesicles or droplets \cite{Tsai2011, Shah2014}. These experimental systems are a useful tool for understanding the more complex dynamics of cells. The model in this section makes predictions of interesting active phenomena including symmetry breaking, and droplet deformation, that are relevant to the field of cell mechanics.

\subsection{Model}
\label{sec:abmodel}

We consider a fluid droplet described by an interfacial surface $\Sigma$ separating the contained fluid domain $\Omega_0$ and external fluid domain $\Omega_1$ with viscosities $\eta_0$ and $\eta_1$ respectively. We define a concentration of active matter $c(\theta,\phi,t)$ on the interface $\Sigma$, which alters the droplet surface tension $\gamma$ such that $\gamma = \gamma_0 - \actc \, c - Bc^2/2$. $\gamma_0$ is the bare surface tension, $\actc$ is the activity ($\actc<0$ for contractile) and $B$ is a passive repulsion force. This higher order repulsive term represents passive pressure, similar to that in \cite{Joanny2013}, which parametrises the compressibility of the active fluid on the interface. We denote the effective surface tension $\gamma'_0 =  \gamma_0 - \actc \, c_0 - Bc_0^2/2$, which is the value of $\gamma$ in the stationary state.

The force density on the droplet interface is then: $\vect{F} = \kappa \gamma \vect{\hat{n}} + \rndb{\nabla_s \gamma}\vect{\hat{t}}_i $, where $\vect{\hat{n}} = \vect{\hat{n}}(\theta,\phi,t)$ is the outward surface normal, $\vect{\hat{t}}_i = \vect{\hat{t}}_i(\theta,\phi,t)$ are the orthogonal surface tangent vectors, $\kappa = \divv{\vect{\hat{n}}}$ is the local curvature, and $\nabla_s = (\vect{\hat{t}}_i \cdot \nabla)$ is the surface gradient. It is useful to define the effective activity $\actt=\actc+Bc_0$ which defines the scale of the force $\vect{F}$ for small deviations of the concentration $c$ from $c_0$. Thus, the interface has net contractility for $\actt<0$.

The only forces acting on the system originate at the droplet surface $\Sigma$, with position $\vect{R} = R(\theta,\phi,t)\hat{\vect{e}}_r$ assuming this is single-valued with respect to the angular coordinates ($\theta$,$\phi$). Thus, the resulting force density in the fluid is $\vect{f}^{\rm ext}(r,\theta,\phi,t) = \vect{F} \delta\sqb{r - R(\theta,\phi,t)}$. We ignore inertia taking the low Reynolds' number limit, $Re = 0$, thus the incompressible fluid flow ($\divv \vect{v} = 0$) is described by Stokes' equation $\eta_n \nabla^2\vect{v} + \vect{f}^{\rm ext} - \nabla P = 0$, where $n=0,1$ denotes the domain $\Omega_0$ or $\Omega_1$, $\vect{v}=\vect{v}(r,\theta,\phi,t)$ is the fluid velocity, $\vect{f}^{\rm ext}=\vect{f}^{\rm ext}(r,\theta,\phi,t)$ denotes any external force densities and $P=P(r,\theta,\phi,t)$ is the hydrostatic pressure. We take the limit of a zero-thickness interface and assume flow and stress continuity between the two fluids $\Omega_0$ and $\Omega_1$. This means the active particles act as an active surfactant, rather than a thin viscous layer (as in \cite{Bois2011, Hawkins2011, Joanny2013, Khoromskaia2015, Turlier2014, Callan-Jones2016}), which allows us to study the dynamics of deformation in a 3D viscous environment analytically.

The evolution of the surface concentration $c$ with respect to time $t$ is:
\begin{eqnarray}
\label{dcdt} \dot{c} = -\nabla_s \cdot (c \vect{v}_b) + D\nabla_s^2 c - k_{\rm off}c + k_{\rm on}\rho_b \,,
\end{eqnarray} 
where $\dot{c}=\partial c/\partial t$, $\vect{v}_b = \vect{v}(r=R,\theta,\phi,t)$ is the interface flow velocity, $D$ is the diffusion constant for the active particles on $\Sigma$, and $k_{\rm on,off}$ are binding and unbinding rates of the particles to the interface. The concentration of unbound particles in the bulk of the drop is denoted $\rho = \rho(r,\theta,\phi,t)$. Binding occurs at the interface where we denote the concentration of unbound prticles $\rho_b = \rho(r=R,\theta,\phi,t)$. Note that $k_{\rm on}$ has units of velocity, as it contains the adsorption depth parameter. We assume that the active particles are insoluble in the external fluid, and so the evolution of the bulk concentration $\rho$ is given by:
\begin{eqnarray}
\label{drhodt} \dot{\rho} = -(\vect{v}\cdot\nabla)\rho + D_\rho\nabla^2 \rho 
\end{eqnarray}
with the boundary condition $D_\rho(\vect{n}\cdot\nabla)\rho = k_{\rm on}\rho - k_{\rm off}c$ at $r=R$, to ensure conservation of mass. The parameter $D_\rho$ is the bulk diffusion constant of the active particles. Here we assume that the active particles only generate stresses at the interface, so the bulk concentration acts as a buffer to recycle the surface concentration. 

\subsection{Linear Stability Analysis}
\label{sec:ablsa}

In this section we present the results of a linear perturbation to the stationary ground state of the droplet. The system is in a stationary (velocity $\vect{v} = 0$) steady state when the interface is spherical (fixed radius $R = R_0$) with a homogeneous concentration of active particles ($c = c_0$). Then the bulk concentration is $\rho_0 = k_{\rm off} c_0/k_{\rm on}$ inside the drop, and the hydrostatic pressure inside is $\vect{P} = P_{\rm ext} + (2\gamma_0 - \actt c_0)/R_0$ where $P_{\rm ext}$ is the stationary state pressure in the external fluid. We perform a linear stability analysis by applying a small perturbation to the variables defined at the interface $R$ and $c$ of the form: $\tilde{g} = g_0 + \sum_{l=1}^{\infty}\sum_{m=-l}^{l}\delta g_{lm}(t)Y_l^m(\theta,\phi)$, where $Y_l^m$ are the spherical harmonic functions and $\delta g_{lm} \ll g_0$. To first order, the resulting flow is given by Lamb's solutions for flow around a sphere, which can be expressed as vector spherical harmonics \cite{Lamb1945}. Solving the Stokes equation with flow and stress continuity conditions at the droplet interface gives expressions for $\delta v^{(i)}_{lm}$ (as defined in \cite{Carrascal1991} and Supplementary Information appendix A) in terms of $\delta c_{lm}$ and $\delta R_{lm}$. The perturbation on the interface is also coupled to a perturbation of the internal concentration $\rho$ such that
\begin{eqnarray*}
\rho = \sqb{\frac{k_{\rm off} c_0}{k_{\rm on}} + \sum_{l=1}^{\infty}\sum_{m=-1}^{l}\delta\rho(r,t)Y_l^m} \, .
\end{eqnarray*}
We obtain analytical solutions for the stability by assuming a quasistatic solution for $\delta \rho$ (taking $\dot{\rho}=0$). This assumption corresponds to a fast relaxation of the bulk concentration $\rho$ compared to the timescale of evolution of the surface concentration $c$. At linear order, the solution for $\delta \rho$ simply satisfies the diffusion equation with a flux condition at the boundary:
\begin{eqnarray*}
\delta \rho = \frac{k_{\rm off} R_0 \delta c}{D_\rho l + k_{\rm on} R_0} \rndb{\frac{r}{R_0}}^l  \,.
\end{eqnarray*}
This solution enables us to predict the effect of the feedback by diffusion through the bulk analytically. The full solutions to the coupled linear equations are solved exactly with Bessel functions as in \cite{Hawkins2011}, however these solutions do not permit an analytical calculation of the stability condition, hence we do not consider them here, but instead compare our approximate analytical solutions directly with the full dynamical simulations.

Finally, we evaluate the coupled system of dynamic equations for the concentration (equation \eqref{dcdt} in section \ref{sec:abmodel}) and radius $\dot{R} = \vect{v}_b.\hat{\vect{n}}$ (the normal velocity at the interface) to first order in the perturbations. We find instabilities by looking for positive eigenvalues of the stability matrix that relates $\dot{c}$ and $\dot{R}$ to $\delta c$ and $\delta R$ to first order in the perturbations (see Supplementary Information appendix A for further details of this calculation). From this analysis we find an instability threshold for the effective activity $\actt<\alpha_I$ where 
\begin{eqnarray}
\label{intZ} \alpha_I = -\frac{2\tilde{\eta}}{c_0}\rndb{2l+1}\rndb{\frac{D}{R_0} + \frac{D_\rho R_0 k_{\rm off}}{(l+1)\rndb{D_\rho l + k_{\rm on} R_0}}},
\end{eqnarray}
where $\tilde{\eta} = (\eta_0 + \eta_1)/2$ is the mean viscosity of the internal and external fluid. We see that $\alpha_I$ is independent of the effective surface tension $\gamma_0'$ which shows that the coupled droplet deformation does not contribute to the symmetry breaking threshold. However, the corresponding maximum eigenvalue of the stability matrix does weakly depend on the effective surface tension $\gamma_0'$ for $l>1$. This weak positive relation suggests that the instability should evolve more quickly in large surface tension drops when $l>1$. In this linear limit there is no contribution from the advection term in \eqref{drhodt} and the second term in \eqref{intZ} (proportional to the binding rates) always increases the threshold. This is because the binding terms allows the concentration on the interface to be recycled by unbinding and diffusing into the bulk of the drop.

{The stability analysis shows how the droplet will initially deform. This deformation is characterised at short times by the maximally unstable mode $l_{\rm max}$, which can be found exactly when binding is not included (see figure \ref{fig:lmax} and Supplementary Information appendix A). At linear order the instability is independent of the spherical harmonic parameter $m$. Generically, $l_{\rm max}$ predicts that as contractile activity is increased, the more concentration peaks will be initially formed on the droplet surface (figure \ref{fig:lmax}). The total droplet activity scales with droplet size, and so $l_{\rm max}$ is more sensitive to the activity parameter $\actt$ in larger droplets. Thus it is easier to observe modes with small $l$ in smaller droplets, where the dynamics are less sensitive to small changes in the activity. Note that only the $l=1$ mode ($k=1$ in 2D) produces net propulsion of the droplet (i.e. $\int_\Sigma \dot{R}\hat{\vect{n}}{\rm d}S \neq 0 $), so the first unstable mode corresponds to front-back symmetry breaking of the droplet profile. 

As shown in Supplementary Information appendix A, one can approximate the maximally unstable mode $l_{\rm max}$ analytically by solving $\dot{R} = 0$ for $\delta R_{lm}$. This approximation imposes that $R$ always assumes the steady state shape for a given fixed concentration perturbation $\delta c_{lm}$ (plotted in figure \ref{fig:lmax}). Physically, this assumes that the shape dynamics are much faster than the concentration dynamics, and so can be taken to be quasistatic. Interestingly, while this assumption does not represent the full coupled dynamics of $\delta c_{lm}$ and $\delta R_{lm}$, it does reproduce the critical activity threshold, and also approximates the mode structure well.

When binding is included ($k_{\rm off} \neq 0$) the dispersion relation changes, and as we see from \eqref{intZ} the active threshold is non-linear in $l$, and hence higher (non-swimming) modes can have lower activity thresholds than the $l=1$ (swimming) mode.} 
\begin{figure}
\hfill\includegraphics[width=\columnwidth]{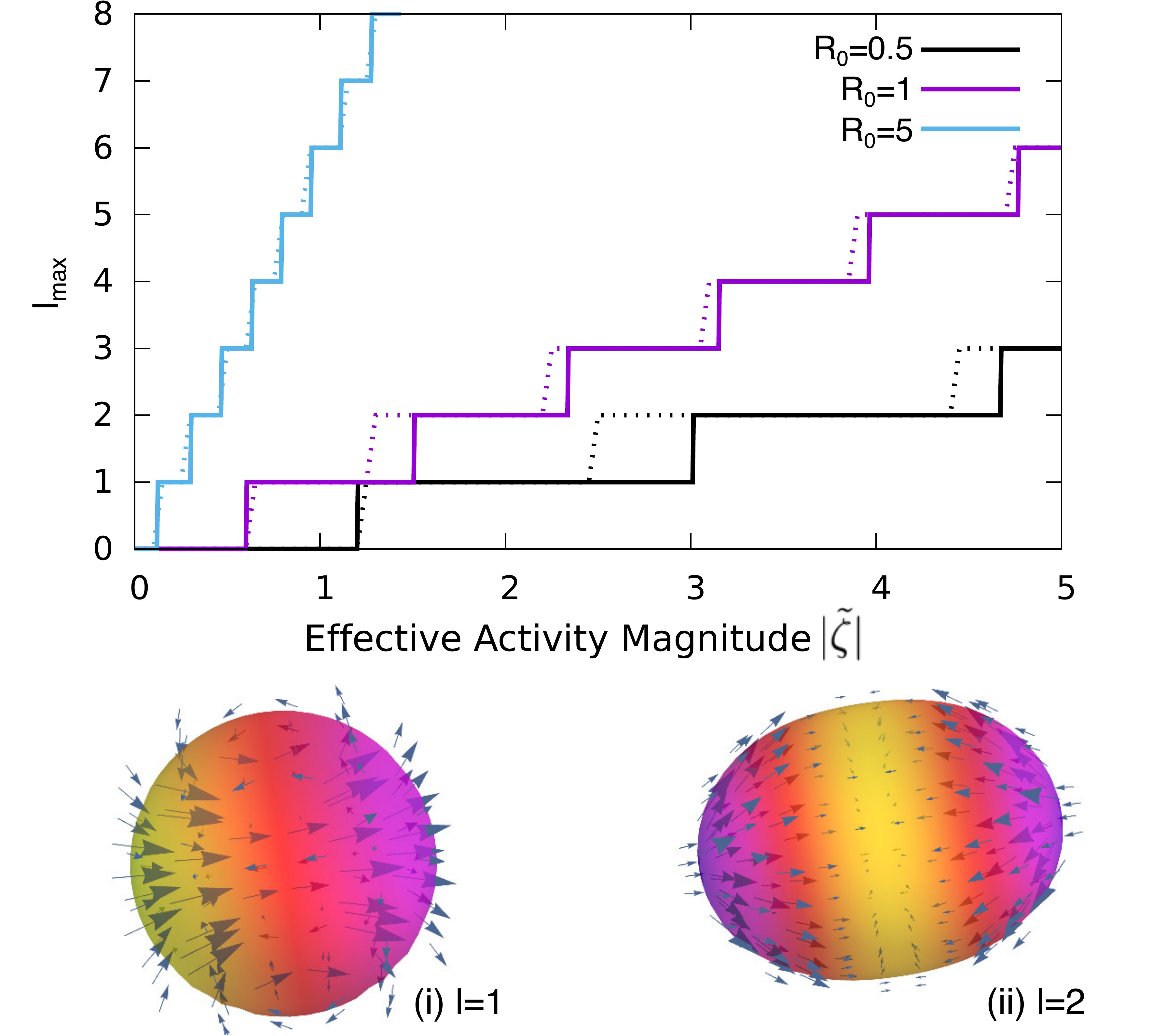}
\caption{Maximum mode number $l_{\rm max}$ plotted against activity in normalised units for increasing values of the droplet radius. Dashed lines show numerical solution and solid lines show analytical approximation using $\dot{R}=0$. Parameters used: $c_0 = 1$, $\gamma_0=1$, $D=0.05$, $\eta_0=\eta_1=1$ and $k_{\rm off}=0$. Insets show flow (blue arrows) and active concentration $c$ (colour gradient  from purple (low) to yellow (high)) to linear order on the perturbed interface for a (i) $l=1$ mode and (ii) $l=2$ mode respectively. Deformation of the interface in (ii) is calculated {by solving $\dot{R}=0$ for $\delta R$ given the form of $\delta c$, and is exaggerated for visibility} using small $\gamma_0'$.}
\label{fig:lmax}
\end{figure}

Within the assumptions made here, the binding and unbinding dynamics always increase the activity threshold. We see that if the binding is fast $k_{\rm on} \gg D_\rho$, the critical activity takes the same form as the 1D model considered in \cite{Hawkins2011} where the active threshold is always minimal for $l=1$ and is proportional to the effective diffusion parameter $\tilde{D} = (Dk_{\rm on} + D_\rho k_{\rm off})/k_{\rm on}$. However, for fast bulk diffusion, geometrical effects become important. A single peak in the interfacial concentration gives rise to a concentration gradient in the bulk driving diffusion away from it. As the number of peaks on the interface increases the concentration gradients are more localised to the surface, and diffusion has a smaller effect. In this regime, the minimum critical activity can correspond to multi-peak modes ($l>1$) when the contribution from bulk diffusion is significant. This is analogous to the findings in \cite{Bois2011} for a one-dimensional active fluid consisting of two-components.

The droplet shape instability is enslaved to the concentration (as  $\alpha_I$ is independent of $\gamma$), so we can estimate how the shape will deform due to certain concentration distributions on the interface by solving $\dot{R}=0$ for $\delta R$ (for $l>1$). Plotted in figure \ref{fig:lmax} is an example of these deformations and the associated flow to linear order. In order to calculate the resulting steady state dynamics we require numerical simulation.}

\subsection{Results and Comparison with Simulations}
\label{sec:absims}

We test these analytical results against the 2D simulations developed in \cite{Whitfield2016}. These use an Immersed Boundary method \cite{Peskin2002,Lai2008} to represent the active interface explicitly as a Lagrangian mesh which is coupled to the Cartesian mesh for the 2D fluid via a numerical Dirac delta function.

\begin{figure}[h!]
	\hfill\includegraphics[width=\columnwidth]{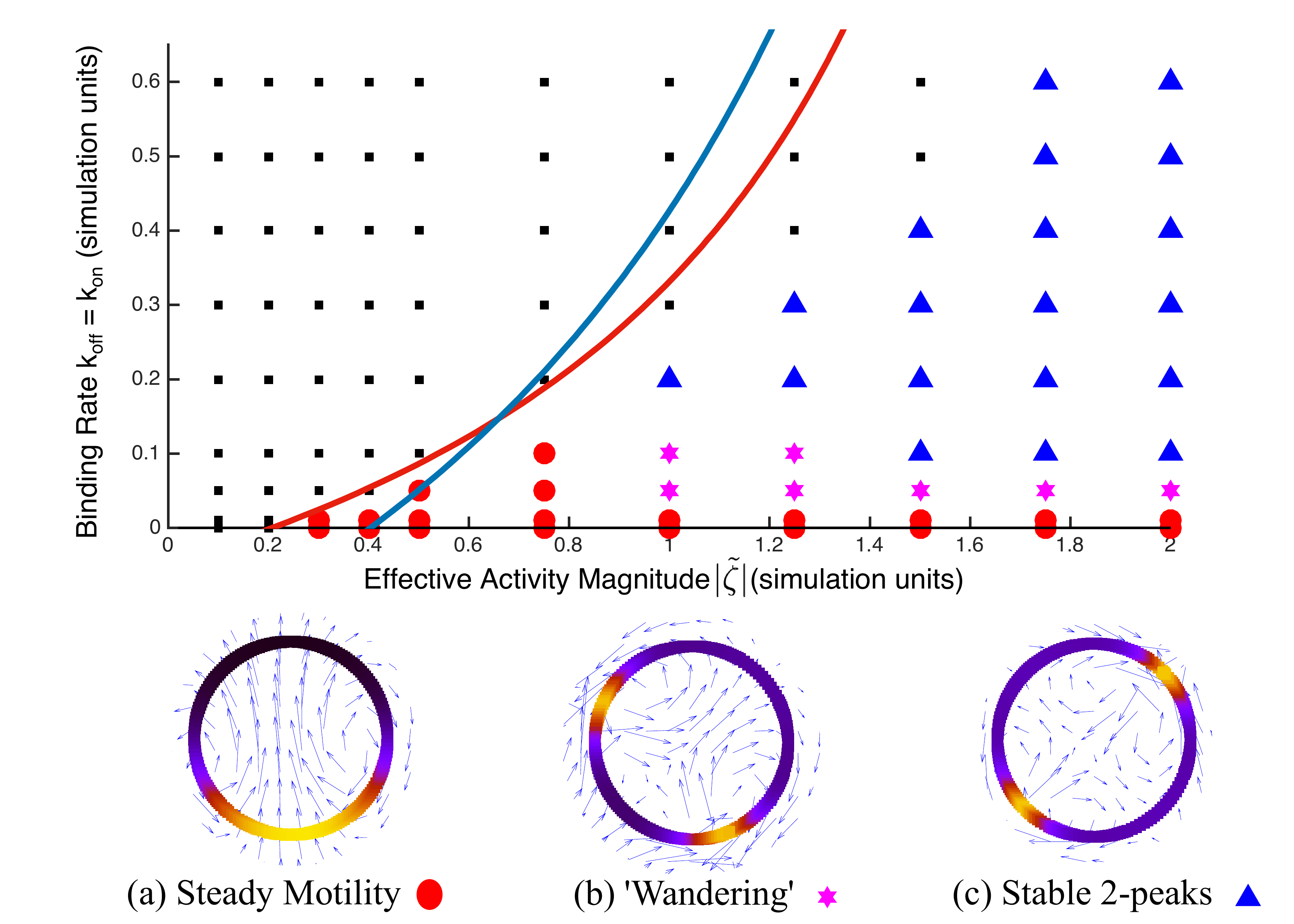}
	\caption{Phase diagram of 2D simulation results for an active isotropic interface, each dot represents a single simulation run. Insets show steady state flow (blue arrows) and concentration fields (colour density, black to yellow) for the different phases. Low values of $k_{\rm off}$ transition from stationary (black squares) to motile (red circles) with a single peak in concentration (shown in (a)). Feedback from the internal concentration {produces intermediate oscillatory states} (magenta stars) and a stationary 2-peak state (blue triangles). Solid lines of increasing gradient show predicted activity threshold for modes $k=1,2$ (red, blue). Simulation parameters: $c_0 = 1$, $R_0=1$, $\gamma_0=1$, $D=0.05$, $D_\rho=0.5$, $\eta_0=\eta_1=1$. }
	\label{fig:abphase}
\end{figure}

Repeating the stability analysis in 2D, we now take perturbations of the form $g = g_0 + \sum_{k=1}^{\infty}{\rm e}^{ik\theta}$. The calculation reveals that surface tension gradients do not deform the drop in 2D (as found in \cite{Yoshinaga2014}) however the concentration dynamics remain very similar. We compare our predictions in 2D to the results of the Immersed Boundary simulations in figure \ref{fig:abphase}. We run simulations varying the activity, binding rate (taking $k_{\rm off} = k_{\rm on}$) and diffusion parameters. At zero binding we observe two steady phases, a stationary state and a steady moving state \ref{fig:abphase}(a) separated by the threshold $\alpha_{\rm I,2D}$ which agrees well with the expected analytical result
\begin{eqnarray}
\label{2DintZ} \alpha_{\rm I,2D} = -4\frac{\tilde{\eta}}{c_0}\rndb{\frac{Dk}{R_0} + \frac{D_\rho R_0 k_{\rm off}}{\rndb{D_\rho k + k_{\rm on} R_0}}} \, .
\end{eqnarray}
{This moving steady state due to a surface tension gradient is also observed for the the self-propelled droplets studied in \cite{Yoshinaga2014,Ohta2009a}. The equations of motion we use (see Model section) are similar to those for the self-propelled droplets studied in \cite{Yoshinaga2014,Ohta2009a} and hence some of the same dynamical behaviour is observed. However, our model predicts new stable states and instabilities corresponding to pure deformation and division as discussed below. This arises due to the advection and diffusion of active particles through the bulk of the drop. Unlike in \cite{Yoshinaga2014,Ohta2009a} the model here conserves the active particles within the drop making it more relevant to cell cortex dynamics.}

We next calculate the maximum mode number $k_{\rm max}$ (see Supplementary Information appendix A). In the regime where we predict $k_{\rm max}=2$, our simulations show initial formation of 2 peaks in droplet concentration. Without binding, these peaks are unstable and always coalesce to form one (as predicted for a flat active viscous layer in \cite{Bois2011}). In this case, the droplet swims persistently and steadily with the concentration peak at its rear. A decomposition of the Fourier modes of this steady state shows that the far field flow is puller like, i.e. its dipole moment is such that it pulls the surrounding fluid inward and pushes it outward along the axis perpendicular to its motion. The activity threshold predicted compares well to that in the simulations for small values of the binding. At larger binding rate, the interior dynamics is not completely diffusion dominated, and the critical activity is underestimated due to the approximation of $\dot{\rho} = 0$. As we increase $k_{\rm off}$ and $\actt$ we see that eventually the droplet becomes immobile with 2 stable peaks in the concentration  (see figure \ref{fig:abphase}). In the intermediate regime the droplet undergoes a `wandering' motion as the concentration profile oscillates between a single peak and two peaks. Equation \eqref{2DintZ} predicts a non-trivial $k$ dependence of the active threshold as binding terms become important. For the parameters used in figure \ref{fig:abphase}, this can be seen by the crossing of the lines for the $k=1$ and $k=2$ modes, meaning that the minimum critical activity is not necessarily for the lowest $k$ mode ($k=1$). Note this is very similar to the prediction in 3D in \eqref{intZ}.

The simulation results in figure \ref{fig:abphase} demonstrate that as the binding rate increases, advection of the concentration through the droplet bulk becomes more important. The advection can stabilise the two peaks at diametrically opposite points on the circle, resulting in a stationary droplet. However, we see that in 2D the drop does not deform, as the radial forces from the activity gradients are always cancelled by the hydrostatic pressure $P$. This is not the case for the full 3D system where we expect concentration gradients to deform the droplets as shown in figure \ref{fig:lmax}. Nonetheless, the 2D simulations show that advection can stabilise the 2 peak configuration, which in 3D would result in symmetric deformation and potentially division of the droplet. Such a 3D simulation is beyond the scope of this work, but would be useful for quantifying the full 3D morphology. Recent work has shown that non-adherent cells exhibit a swimming state similar to the motion described here, and so it would be of interest to test in future work whether the steady state shape in 3D for the model here resembles the `pear shape' observed in \cite{Ruprecht2015, Callan-Jones2016}.

\section{Active Polar Fluid Droplet}

In this section, we consider a droplet filled with an active polar liquid crystal of constant density everywhere. Realising this system experimentally in droplet systems requires high concentrations of active material so that the polar to isotropic phase transition is localised to the droplet centre. This has been achieved \emph{in vitro} for microtubule based active nematics but only in thin films thus far \cite{Sanchez2012,Keber2014}. In these systems the measured order parameter is approximately constant everywhere except in the vicinity of topological defects. Thus we consider the limit where the active fluid is strongly polarised and restrict the analysis to only the orientational degrees of freedom of the active liquid crystal, and do not consider the density or polarisation magnitude degrees of freedom.

\subsection{Model}
\label{sec:apfmodel}

We utilise the model of an active polar fluid developed by Kruse et al. in \cite{Kruse2004,Kruse2005,Furthauer2012} which has similarities to other continuum models of the cytoskeleton on surfaces (such as \cite{Lober2015, Ziebert2014}). We consider the case where the active fluid has strong local ordering and is far from the isotropic phase so that $\norm{\vect{p}}=1$ everywhere (except at defects where $\vect{p}$ is undefined). This approximation is commonly used to model active and passive liquid crystal systems analytically.

In the $Re=0$ limit the total stress in the active polar fluid, $\sigma_{i j}^{\mathrm{tot}} = \sigma_{ij}^{\mathrm{visc}} + \sigma_{ij}^{\mathrm{dist}} + \sigma_{ij}^{\mathrm{act}} $,
has viscous, distortion and active contributions respectively where:
\begin{align*}
\sigma_{ij}^{\mathrm{visc}} &= 2\eta_{n} u_{ij} = \eta_{0,1}\rndb{\partial_i v_j + \partial_j v_i} \, , \\
\sigma_{ij}^{\mathrm{dist}} &= \frac{\nu}{2} \rndb{p_i h_j + p_j h_i} + \frac{1}{2}\rndb{p_i h_j - p_j h_i} + \sigma_{ij}^{\mathrm{e}} \, , \\
\sigma_{ij}^{\mathrm{act}} &= - \act p_i p_j \, .
\end{align*}
The viscous stress is the response to flow assuming a Newtonian fluid. The distortion stress is that of a passive polar liquid crystal due to deviations in filament alignment, where the perpendicular part of the molecular field $h_i = -(\delta F/\delta p_j)(\delta_{ij}-p_ip_j)$ acts to minimise the free energy functional $F = \int_{\Omega+\Sigma} {\rm d}^3r f$ with respect to $\vect{p}$, given $\left| \vect{p} \right| = 1$. The Ericksen stress, $\sigma_{ij}^{\mathrm{e}} = f\delta_{ij} - (\partial f/(\partial(\partial_jp_n)))(\delta_{ij} - p_n p_k) \partial_i p_k $, is a generalisation of the hydrostatic pressure for complex fluids. Finally, the active stress represents the active dipolar force and thus is second order in $\vect{p}$.

The free energy functional $F$ gives the equilibrium properties of the system. Here for simplicity we use the one constant approximation of the Frank free energy:
\begin{eqnarray}
\label{FreeE} F = \int_\Omega {\rm d}^3r \frac{K}{2}(\partial_i p_j)^2 + \int_\Sigma {\rm d}S f_s \; ,
\end{eqnarray}
where $K$ is the elastic constant and $\norm{\vect{p}}=1$. Since we are modelling a finite droplet, the surface terms are important. We consider normal anchoring of the filaments to the interface, with surface distortion free energy density $f_s = W (\vect{p}\cdot\hat{\vect{n}} - 1)^2 $. {This form of the surface free energy includes the `spontaneous splay' term which is allowed in polar liquid crystals \cite{Pleiner1989}.}

The polarisation flux is 
\begin{eqnarray}
\label{dpdt} \dot{\vect{p}} = -\rndb{\vect{v}\cdot\nabla}\vect{p} - \tens{\omega}\cdot\vect{p} - \nu\tens{u}\cdot\vect{p} + \frac{\vect{h}}{\Gamma}
\end{eqnarray}
where $\omega_{ij} = (\partial_iv_j - \partial_jv_i)/2$ and $\Gamma$ is the rotational viscosity.

\subsection{Linear Stability Analysis}
\label{sec:apflsa}

We contrast the model of an active interface to that of a droplet of active polar fluid of constant density. In this case, rather than the concentration of active particles, the important degree of freedom is the polarisation vector $\vect{p}$ denoting the average direction of the contractile filaments in the fluid. 

We calculate the linear stability of the droplet in the limit of strong anchoring $W \rightarrow \infty$ in order to study the effects between the coupling of droplet morphology and polarisation. This equates to the boundary condition $\vect{p} = \hat{\vect{n}}$ at $\vect{r}=\vect{R}$. In the case of weak or no anchoring, instabilities can occur for both extensile ($\act>0$) and contractile ($\act<0$) active polar drops as shown analytically in \cite{Whitfield2015} and in simulations \cite{Tjhung2012}. The condition of fixed polarisation at the interface inhibits certain deformations of the polarisation field at low activities and so the preferred deformation modes are those which can couple to the droplet deformation. This was demonstrated in 2D simulations of active nematic drops in \cite{Giomi2014}. Here we explain this mechanism analytically in a 3D fluid drop by linear stability analysis. The polar nature of the anchoring produces a ``radial hedgehog'' topological defect at the droplet centre (or a radial defect with $+1$ winding number in 2D), giving a simple analytical description of the stationary state. Thus we are able to make analytical predictions about spontaneous symmetry breaking in these systems even in the general 3D case.

Unlike the case of an active interface, the active fluid here fills the drop, and hence active and passive stresses are generated in the bulk. The stationary steady state is given by the polarisation $\vect{p} = \hat{\vect{r}}$, $\vect{R} = R_0 \hat{\vect{r}}$, and $\vect{v} = 0$.

{To perform a general linear stability analysis, one would need to consider generic perturbations to both the polarisation field and interface and study the coupled equations for their evolution, this is not analytically tractable in this case. However, we can perform restricted perturbations that we expect to be representative of the dynamics in a particular limit. We consider the case where the polarisation field is enslaved everywhere to the shape of the boundary by the anchoring condition. This corresponds to the limit where bulk instabilities in the droplet are suppressed by its size (\emph{i.e.} small droplets). In larger droplets, (or equivalently for smaller $K$) the dynamics of the polarisation field becomes more independent of the anchoring condition, and we expect this approximation to break down. }

Due to the symmetry of the stationary state, we first need to consider the special case of the translational mode of perturbation, corresponding to the $l=1$ spherical harmonic mode. Without loss of generality we consider a perturbation along the $z$-direction ($m=0$). This mode implies a translation of the hedgehog defect away from the droplet centre. If we assume that the defect has some fixed finite core radius $R_{c}$ then we can treat the liquid crystal as contained between two boundary conditions, one at the defect $r=R_{c}$ and one at the droplet interface $r=R_0-\delta z \cos(\theta)$, where $\delta z$ is a small displacement of the defect position from the droplet centre along the $z$-direction. The calculation is done in the reference frame of the defect so that it coincides with the origin of our coordinate system. In the equilibrium case ($\act=0$), we can write a polarisation field to first order that minimises the bulk free energy in \eqref{FreeE} by solving $\vect{h} = 0$ for these boundary conditions:
\begin{eqnarray}
\label{pl1} \vect{p}_{l=1} = \vect{e}_r - \delta z \frac{r-R_c}{r(R_0-R_{c})}\sin(\theta) \vect{e}_\theta  \, .
\end{eqnarray}
This method equates the defect to a small colloid with (polar) homeotropic anchoring, and in the strong anchoring case we expect the free energy minimum to correspond to the defect being positioned at the droplet centre as we observe in simulations, and is reported in \cite{Lubensky1998, Poulin1998}. Using the polarisation in equation \eqref{pl1} we can estimate what the bulk free energy increase will be for such a deformation (details in Supplementary Information Appendix B)
\begin{align}
\notag \Delta F_{\rm bulk} &= \frac{4  K \pi \delta z^2}{3 R_{0} (1-\epsilon)^2} \sqb{4- 3\epsilon - \epsilon^2 + 4\epsilon \log \rndb{\epsilon}} +\,  O(\delta z^3) \\
\label{deltaF} &\approx \frac{16  K \pi \delta z^2}{3R_{0}}
\end{align}
where $\epsilon=R_{c0}/R_{0}$ is assumed small in the final approximation of the equation. This $\Delta F$ is positive for all $\epsilon$, suggesting that the free energy minimum corresponds to the defect being positioned at the droplet centre. Note that this polarisation field is only valid to first order in $\delta z$ and so higher order terms could affect the form of the quadratic term here. 

\begin{figure}[h!]
\hfill\includegraphics[width=\columnwidth]{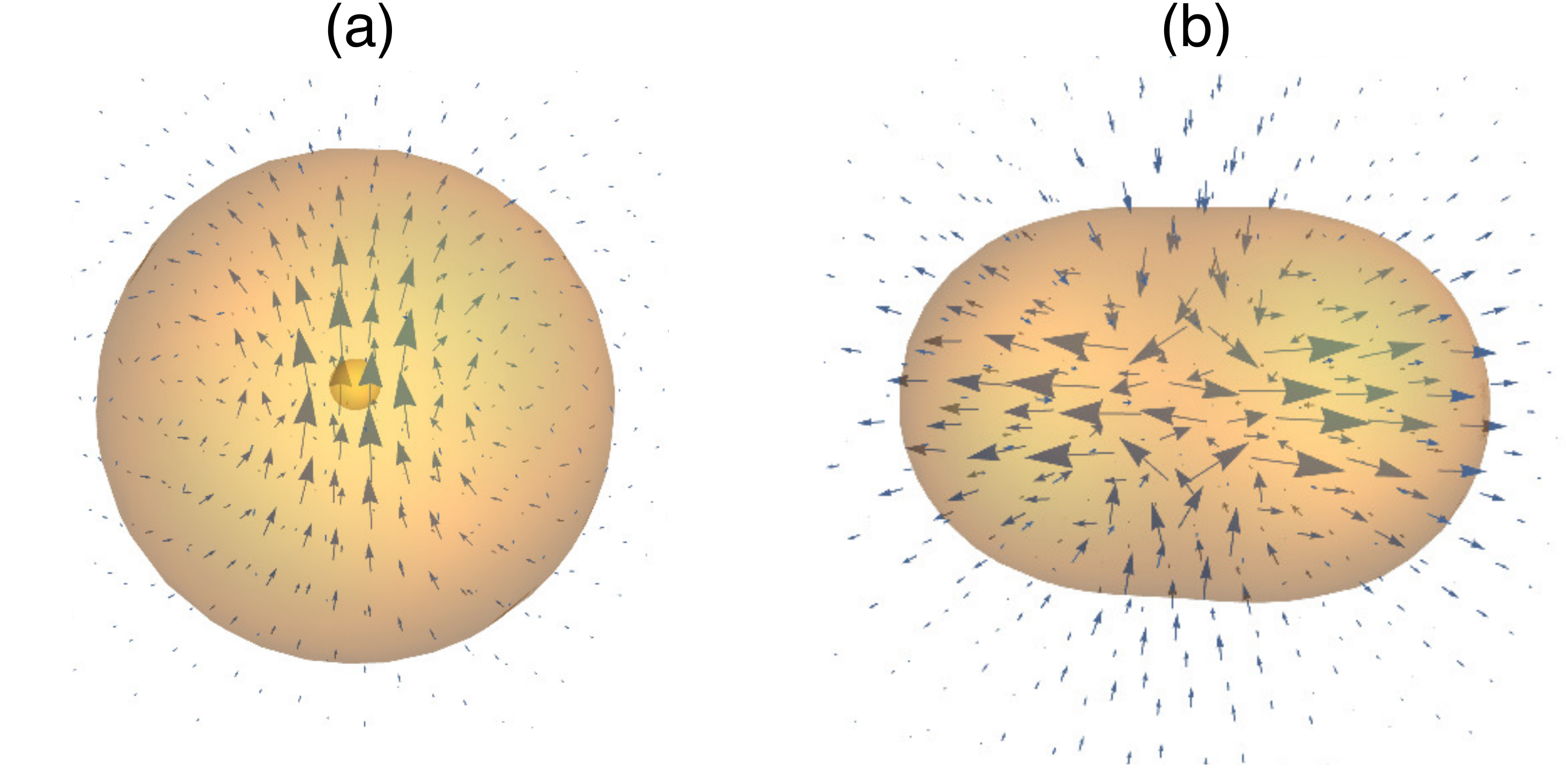}
\caption{Active part of the flow field (blue arrows) to linear order in the perturbations for: (a) defect position (inner sphere) displaced in the vertical direction with $\act>0$ (extensile activity); (b) $l=2$ mode perturbation of the interface assuming strong anchoring of the polarisation field with $\act<0$ (contractile activity). The perturbations are made artificially large for visibility here.}
\label{fig:lsavel}
\end{figure}

We now introduce a small activity $\act$, such that equation \eqref{pl1} remains a valid approximation for the form of the polarisation field, then we see that this gives rise to active forces in the drop. We solve the force balance equations (omitting passive contributions, see Supplementary Information Appendix B) to find the active contribution to the flow. We then integrate to find the active contribution to the velocity of the defect core $\vect{v}_{c}$ and droplet $\vect{v}_{\rm drop}$. The relative velocity of the defect is then:
\begin{equation}
\Delta \vect{v} \equiv \vect{v}_c - \vect{v}_{\rm drop} \approx \act \delta z\frac{(2\eta_0+\eta_1)-\epsilon(\eta_0+\eta_1)}{2 \eta_0 (3 \eta_0 + 2 \eta_1)} \hat{\vect{e}}_z \; .
\end{equation}
We see that extensile activity ($\act>0$) always results in a relative defect velocity that is in the same direction as the initial defect displacement (along $\hat{\vect{e}}_z$), as shown by figure \ref{fig:lsavel}. This implies that extensile activity will destabilise the defect from the centre and give rise to motion of the droplet as a whole (which to linear order is also along $\hat{\vect{e}}_z$). Conversely, we expect contractile activity to stabilise the defect at the droplet centre, as the flows resulting from contractile activity ($\act<0$) act to restore the defect back to its stationary position at the droplet centre.

Thus, within the assumptions made above, one can predict that the active polar droplet will break translational symmetry spontaneously above some finite activity. This mode of symmetry breaking is independent of surface deformations at linear order, and so its critical activity threshold should not depend on the droplet surface tension. Hence the critical activity threshold will only depend on the increase in the passive free energy (equation \eqref{deltaF}), which goes to a finite value in the limit of a point defect and scales as the inverse of the droplet size. In general, the parameter $\epsilon$ is difficult to define, which is a consequence of the  assumption of $\norm{\vect{p}}=1$, which breaks down around the defect. This can be avoided by using a Landau-De Gennes type free energy description for the passive part of the dynamics such that there is an polar-to-nematic phase transition at the centre of the droplet. However, such an approach is not analytically tractable, as it requires solving non-linear partial differential equations for the radial dependence of $\vect{p}$. Qualitatively though, the predictions here are consistent with what is observed in the simulations.

For perturbation modes $l>1$ the flow at the origin will always be zero, and so one can assume that in the strong anchoring limit the defect will remain centred at the origin. We again require an assumption for the $r$-dependence of the polarisation perturbation. Taking $R_{c0} \rightarrow 0$, we can write a general form as $\delta \vect{p} \propto r^n$ for arbitrary $n\ge0$. Importantly, for all $n$, the active flows always give rise to an instability for $\act<0$ (contractile). Considering only active flows, the maximally unstable perturbation is for $n=0$. Thus, below we consider only the results of this mode, which allows us to consider the dynamics in the limit where the filament polarisation at the interface and in the droplet are strongly coupled. However it comes at the cost of reducing the quantitative power of our predictions, and is an important restriction to the dynamics considered. Note, in 2-dimensions, the assumption $n=0$ gives rise to an infinite passive contribution to the dynamics (proportional to $K$) and so we use $n=1$, which appears consistent with what is observed in simultions.

In the strong anchoring limit, the polarisation has to match the perturbed interface normal at $r=R$ to first order, such that
\begin{eqnarray}
\label{pl2} \vect{p} = \hat{\vect{r}} - \sum_{l=2}^\infty \sum_{m=-l}^{l} &\sqb{\frac{\delta R_{lm}(t)}{R_0} r(\nabla Y_l^m(\theta,\phi))} \, .
\end{eqnarray}
We calculate the resulting flows to first order in $\delta R$. Since $\vect{p}$ is enslaved to the deformation we then only need to consider the radius dynamics given by $\dot{R}$ (for details see Supplementary Information appendix B).

In this strong anchoring limit we find that the droplet is unstable if $\act<\alpha_P<0$, i.e. the activity threshold, $\alpha_P$, is always contractile. The threshold $\alpha_P$ increases linearly with $\gamma$ and $K$. Repeating the linear stability analysis calculation in 2D shows the same qualitative prediction, where this time we take $\delta p \propto r$ as this is the leading order contribution allowed. The analytical expressions for the activity threshold are given in Supplementary Information appendix B and a full discussion of the eigenvalues of the general stability matrix (for weak anchoring) can be found in \cite{Whitfield2015}.

The result of this analysis is somewhat surprising, in this strong anchoring limit we expect the $l=1$ mode to be unstable to extensile activity, whereas the higher modes of deformation are unstable for contractile activity. This suggests that, when our assumptions hold, we should see translational symmetry breaking with the defect moving to the droplet front for an extensile drop and symmetric modes of deformation for a contractile drop (see figure \ref{fig:lsavel}). This active threshold scales linearly with $K$ and $\gamma_0$, demonstrating the importance of the coupling of the morphology to the polarisation field. Contrast this to the case of the active interface where the shape does not affect the threshold for a phase transition.

\begin{figure}[h]
\centering
\hspace{30mm}\includegraphics[width=\columnwidth]{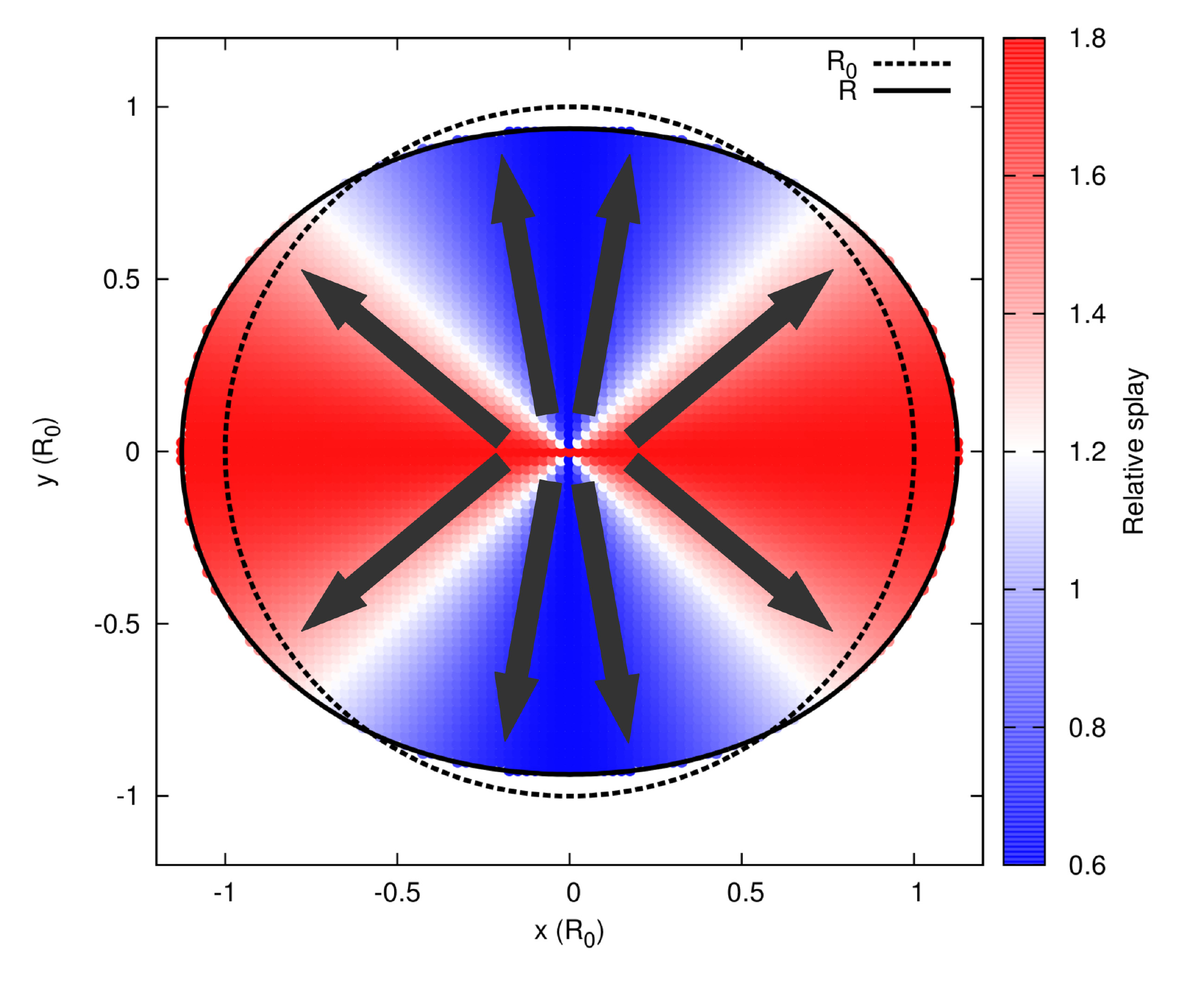}
\caption{Spatial change in splay induced by boundary pertubation. Dotted line indiciates $R_0$ and solid line the perturbed interface $R$. Increased splay in regions of higher curvature drive outward flows, coupling to further increase in boundary curvature. The black arrows indicate polarisation direction while the colour gradient indicates the splay magnitude $\norm{\divv{\vect{p}}}$ relative to its value in the stationary state.}
\label{fig:instsketch}
\end{figure}

This contractile instability can be understood physically by considering the splay in the drop due to perturbations in the interface curvature. High curvature couples to increased splay which couples to outward flow, further increasing the curvature of the interface and hence the splay. A sketch of this is given in figure \ref{fig:instsketch}. 

\subsection{Results and Comparison with Simulations}
\label{sec:apfsims}

\begin{figure}[h!]
	\hfill\includegraphics[width=\columnwidth]{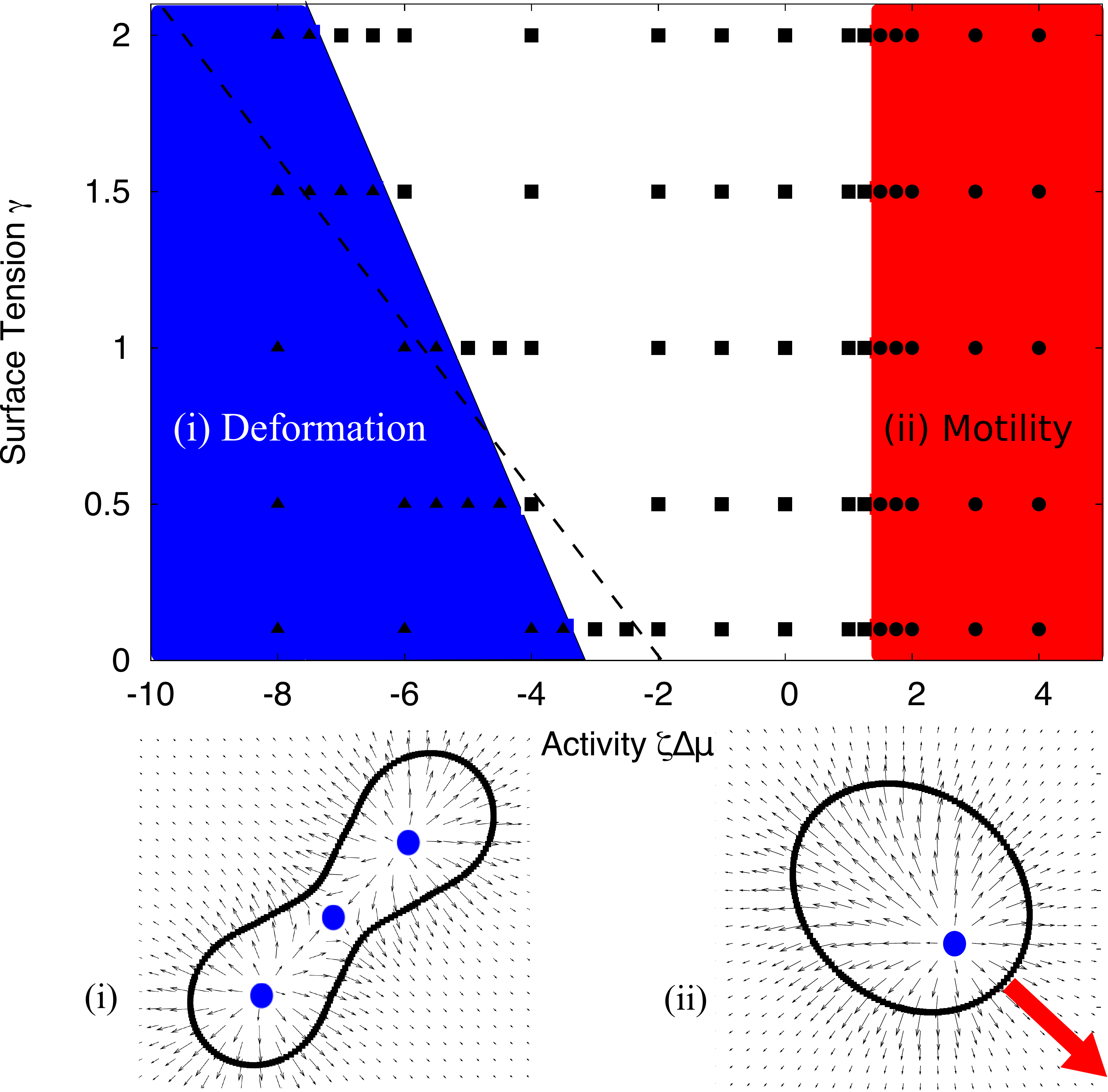}
	\caption{Active polar drop stability diagram. Stationary state (white, square dots), spontaneous symmetric deformation (blue, triangular dots) and spontaneous motility (red, round dots) are observed. Dashed line shows analytical prediction from linear stability analysis. Insets show the polarisation field $\vect{p}$ (black arrows) inside the droplet following symmetry breaking with defects labelled by blue dots. Note that due to the simulation method, the polarisation field in the simulations changes continuously from $\norm{p}=1$ inside the drop to $\norm{p}=0$ outside, hence the polarisation is defined everywhere in (i) and (ii). Parameters used: $K = 0.1$, $R_0=1$, $\eta_0=\eta_1=\Gamma=1$, $W = 50$ and $\nu=1.1$.}
	\label{fig:bulkphase}
\end{figure}

In the 2D simulations (see figure \ref{fig:bulkphase}) we see symmetry breaking corresponding to the $k=1$ mode for extensile activity resulting in a steady motile state, as predicted by the stability analysis. This is characterised by the defect centre moving to the front of the drop and is independent of the boundary deformation (and hence $\gamma_0$). Due to the extensile nature of the activity this droplet is a pusher, pushing fluid out along its axis of motion and thus elongating parallel to its motion.  

Conversely contractile activity stabilises the defect at the droplet centre and we observe a $k=2$ mode instability characterised by deformation of the droplet into a `dumbbell' shape. It is also observed that this phase behaviour breaks down as the value of $K/R_0^2$ is reduced. In this limit the distortions in the droplet bulk are not strongly coupled to those at the interface and so more complex distortions can occur without significant droplet deformation. Our analytical calculations do not predict this as we assume a form for the $r$-dependence of the polarisation such that it is strongly coupled to the curvature.  This behaviour goes beyond the scope of the analytical work here as this corresponds to a transition to an `active turbulence' state, as numerically simulated in \cite{Whitfield2016}. 

Finally, we also observe rotational steady states in the simulations (for extensile activity when using $\nu=-1.1$) which can be characterised exactly by rotationally invariant distortions of the polarisation field \cite{Kruse2004,Kruse2005}, but these are not predicted for the parameter range used in figure \ref{fig:bulkphase}. 

\section{Discussion}
\label{sec:discussion}

We have used analytical linear stability analysis and numerical simulation to characterise instabilities in active droplets and their resulting non-equilibrium steady states. Recent advances in experimental techniques mean that active gels of cytoskeletal material can be produced \emph{in vitro}. The predictions of our active interface model could be tested by adsorbing an isotropic actin gel onto the interface of an emulsion drop containing myosin and ATP \cite{Tsai2011,Shah2014}. We predict an activity threshold for spontaneous motion, and a further continuous transition to a stable symmetric state mediated by advection of motors through the droplet bulk. We predict that in 3D this symmetric configuration will be coupled to deformation of the drop, however this cannot be observed in the 2D model.

The active polar drop model we use only predicts some of the dynamics of a real active polar drop system as it ignores the density and ordering magnitude degrees of freedom. However, this model system gives us an insight into the intrinsic instabilities when droplet deformation and filament polarisation direction are strongly coupled. In particular, there is a contractile activity threshold that is linearly dependent on surface tension, above which the droplet spontaneously deforms into a characteristic dumbbell shape. We also see persistent motility in the case of extensile activity such that the droplet acts as a \emph{pusher}, compared to the \emph{puller} type motion exhibited in the active isotropic interface model. This is consistent with previous active droplet models that show contractile activity resulting in droplets which are \emph{pullers} and extensile activity resulting in \emph{pushers} \cite{Tjhung2012, Giomi2014, Marth2014, Khoromskaia2015, Tjhung2015}. An interesting future extension of this work would be to consider coupling between both of the active phases studied here within a single drop.

The finite active systems we study improve our understanding of how confinement and deformation affect steady state dynamics. Additionally, we see the importance of feedback, driven by advection through the droplet or the internal orientational order, resulting in more complex dynamics. These results should prove useful in characterising future experiments on \emph{in vitro} cytoskeletal networks and be useful in developing more complex models of multicomponent active systems in nature.

\section*{Acknowledgements}

We acknowledge the EPSRC for funding this work, grant reference EP- K503149-1.



\section*{Supplementary material}

# SUPPLEMENTARY INFORMATION: APPENDICES

## Appendix A: Flow field and Stability of Active Interface.

The generic perturbation to the active interface model is of the form:

$$c = c_0 + \sum_{l=1}^{\infty} \sum_{m=-l}^{l} \delta c_{lm} Y_l^m(\theta, \phi), \tag{S.1}$$

$$\boldsymbol{v} = \sum_{l=1}^{\infty} \sum_{m=-l}^{l} \left[ \delta v_{lm}^{(1)}(r) Y_l^m(\theta, \phi) \hat{\boldsymbol{r}} + \delta v_{lm}^{(2)}(r) r (\nabla Y_l^m(\theta, \phi)) + \delta v_{lm}^{(3)}(r) \left( \boldsymbol{r} \times \nabla Y_l^m(\theta, \phi) \right) \right], \tag{S.2}$$

$$\boldsymbol{R} = R \hat{\boldsymbol{e}}_r = \left\{ R_0 + \sum_{l=1}^{\infty} \sum_{m=-l}^{l} \delta R_{lm} Y_l^m(\theta, \phi) \right\} \hat{\boldsymbol{e}}_r, \tag{S.3}$$

$$P = P_0 + \sum_{l=1}^{\infty} \sum_{m=-l}^{l} \delta P_{lm}(r) Y_l^m(\theta, \phi). \tag{S.4}$$

where we assume the perturbations $\delta g_{l,m} \ll 1$ for $g = c, v^{(i)}$, and $R$. Note that all real perturbations will satisfy the constraint $\delta g_{lm} = (-1)^m \delta g_{l,-m}^*$. From equation (??) we can calculate the perturbed normal and tangential vectors to first order in $\delta R_{lm}$:

$$\hat{\boldsymbol{n}} = \hat{\boldsymbol{e}}_r - \sum_{l=1}^{\infty} \sum_{m=-l}^{l} \frac{\delta R_{lm}}{R_0} \left[ \frac{\partial Y_l^m}{\partial \theta} \hat{\boldsymbol{e}}_\theta + \frac{1}{\sin(\theta)} \frac{\partial Y_l^m}{\partial \phi} \hat{\boldsymbol{e}}_\phi \right], \tag{S.5}$$

$$\hat{\boldsymbol{t}}_1 = \left[ \sum_{l=1}^{\infty} \sum_{m=-l}^{l} \frac{\delta R_{lm}}{R_0} \frac{\partial Y_l^m}{\partial \theta} \right] \hat{\boldsymbol{e}}_r + \hat{\boldsymbol{e}}_\theta, \tag{S.6}$$

$$\hat{\boldsymbol{t}}_2 = \left[ \sum_{l=1}^{\infty} \sum_{m=-l}^{l} \frac{\delta R_{lm}}{R_0} \frac{1}{\sin(\theta)} \frac{\partial Y_l^m}{\partial \phi} \right] \hat{\boldsymbol{e}}_r + \hat{\boldsymbol{e}}_\phi. \tag{S.7}$$

Substituting (??) and (??) into the expression for the force on the interface gives:

$$\boldsymbol{F}_{lm} = -\frac{1}{R_0} \left\{ 2\gamma_0' + \left[ \gamma_0'(l+2)(l-1) \frac{\delta R_{lm}}{R_0} - 2 \delta c_{lm} \tilde{\zeta} \right] Y_l^m \right\} \hat{\boldsymbol{e}}_r$$
$$+ \frac{1}{R_0} \left[ 2\gamma_0' \frac{\delta R_{lm}}{R_0} - \tilde{\zeta} \delta c_{lm} \right] \frac{\partial Y_l^m}{\partial \theta} \hat{\boldsymbol{e}}_\theta + \frac{1}{R_0 \sin(\theta)} \left[ 2\gamma_0' \frac{\delta R_{lm}}{R_0} - \tilde{\zeta} \delta c_{lm} \right] \frac{\partial Y_l^m}{\partial \phi} \hat{\boldsymbol{e}}_\phi \tag{S.8}$$

where $\gamma_0' = \gamma_0 - \tilde{\zeta} c_0 - B c_0^2/2$ and $\tilde{\zeta} = \zeta + B c_0$ as defined in the text. As this force is located at the interface, we can solve the fluid flow in the bulk with Lamb's general solutions for flow around a sphere, which defines the $\delta v_{lm}^{(i)}$ and $\delta P_{lm}$ functions:

$$\delta v_{lm}^{(1)}(r) = \begin{cases} a_{lm}^{(0)} r^{l+1} + a_{lm}^{(1)} r^{l-1} & \text{if } r < R \\ b_{lm}^{(0)} r^{-l} + b_{lm}^{(1)} r^{-l-2} & \text{if } r > R \end{cases} \tag{S.9}$$

$$\delta v_{lm}^{(2)}(r) = \begin{cases} \left( a_{lm}^{(0)}(l+3) r^{l+1} + a_{lm}^{(1)}(l+1) r^{l-1} \right) / [l(l+1)] & \text{if } r < R \\ -\left( b_{lm}^{(0)}(l-2) r^{-l} + b_{lm}^{(1)} l r^{-l-2} \right) / [l(l+1)] & \text{if } r > R \end{cases} \tag{S.10}$$

$$\delta v_{lm}^{(3)}(r) = \begin{cases} a_{lm}^{(2)} r^l & \text{if } r < R \\ b_{lm}^{(2)} r^{-l-1} & \text{if } r > R \end{cases} \tag{S.11}$$

$$\delta P_{lm}(r) = \begin{cases} 2\eta_0 a_{lm}^{(0)} r^l (2l+3)/l & \text{if } r < R \\ 2\eta_1 b_{lm}^{(0)} r^{-l-1} (2l-1)/(l+1) & \text{if } r > R. \end{cases} \tag{S.12}$$

We can then find the coefficients $a_{lm}^{(i)}$ and $b_{lm}^{(i)}$ by solving the boundary conditions of flow and stress continuity (or force balance), including the contribution from the active interface:

$$\boldsymbol{v}^{\text{ext}}\big|_{r=R} - \boldsymbol{v}\big|_{r=R} = 0 \tag{S.13}$$

$$2\hat{\boldsymbol{n}} \cdot \left[\eta_0 \underline{\underline{u}} - \eta_1 \underline{\underline{u}}^{\text{ext}}\right]\big|_{r=R} = \boldsymbol{F}_{lm} + (P - P^{\text{ext}})\hat{\boldsymbol{n}}\big|_{r=R} \tag{S.14}$$

where $\underline{\underline{u}} = (\nabla \boldsymbol{v} + (\nabla \boldsymbol{v})^{\text{T}})/2$ is the strain rate tensor and the superscript 'ext' denotes a property of the external fluid $\Omega_1$. Solving these for the flow and pressure from equations (??)-(??) we find:

$$a_{lm}^{(0)} = \frac{l(l+1)(l+2)}{2\eta'}\left[\gamma_0'\frac{\delta R_{lm}}{R_0}(l-1) - \tilde{\zeta}\delta c_{lm}\right]R_0^{-l-1} \tag{S.15}$$

$$b_{lm}^{(0)} = -\frac{l(l+1)(l-1)}{2\eta^*}\left[\gamma_0'\frac{\delta R_{lm}}{R_0}(l+2) + \tilde{\zeta}\delta c_{lm}\right]R_0^l \tag{S.16}$$

$$a_{lm}^{(1)} = -\frac{l(l+1)}{2\eta'\eta^*}\left\{\gamma_0'\frac{\delta R_{lm}}{R_0}(l+2)(l-1)[3\eta_1 + 4\tilde{\eta}l(l+2)] - \tilde{\zeta}\delta c_{lm}\left[6(\eta_1 - \eta_0) + 3l\eta_1 + 4\tilde{\eta}l^2(l+2)\right]\right\}R_0^{-l+1} \tag{S.17}$$

$$b_{lm}^{(1)} = \frac{l(l+1)}{2\eta'\eta^*}\left\{\gamma_0'\frac{\delta R_{lm}}{R_0}(l+2)(l-1)(\eta_0 - 2\eta_1 + 4l^2\tilde{\eta}) + \tilde{\zeta}\delta c_{lm}\left[4\eta_1 - 5\eta_0 + l(\eta_0 - 2\eta_1) + 4l^2(l+1)\tilde{\eta}\right]\right\}R_0^{l+2} \tag{S.18}$$

$$a_{lm}^{(2)} = b_{lm}^{(2)} = 0. \tag{S.19}$$

where $\eta' = [3\eta_0 + 4\tilde{\eta}l(l+2)]$ and $\eta^* = (\eta_1 - 2\eta_0 + 4l^2\tilde{\eta})$, and $\tilde{\eta} = (\eta_0 + \eta_1)/2$ as in the text. In order to satisfy the binding and unbinding dynamics, we consider the limit $\dot{\rho} = 0$, which gives $\delta\rho_{lm}(r,t) = \delta c_{lm}k_{\text{off}}R_0^{(1-l)}r^l/(D_b l + k_{\text{on}}R_0)$. We then use these solutions in the dynamic equations (1) and $\dot{R} = \boldsymbol{V}\cdot\boldsymbol{n}$ which can be written in matrix form as:

$$\frac{d}{dt}\begin{pmatrix}\delta c_{lm}\\ \delta R_{lm}\end{pmatrix} = \underline{\underline{T}}\begin{pmatrix}\delta c_{lm}\\ \delta R_{lm}\end{pmatrix}$$

$$\text{where}\quad \underline{\underline{T}} = \begin{pmatrix} X - \tilde{\zeta}(c_0/R_0)(4\eta_1 - 5\eta_0 + 2l^2(l+3)\tilde{\eta})Y & \frac{\gamma_0'c_0}{R_0^2}(l+2)(l-1)(2\eta_1 - \eta_0 + 2l\tilde{\eta})Y \\ \tilde{\zeta}(2\eta_1 - \eta_0 + 2l\tilde{\eta})Y & -2\frac{\gamma_0'}{R_0}(l+2)(l-1)\tilde{\eta}Y \end{pmatrix} \tag{S.20}$$

where $X = -Dl(l+1)/R_0^2 - D_b k_{\text{off}}l/(D_b l + k_{\text{on}}R_0)$ and $Y = l(l+1)/(\eta'\eta^*)$. Clearly, when $l = 1$, there is no deformation and the eigenvalue $\lambda$ of $\underline{\underline{T}}$ is just equal to $T_{11}$. In general the solution for the eigenvalues is the root of a quadratic. One of the eigenvalues is always zero when $\gamma_0' = 0$, and is positive for $\gamma_0' < 0$. This instability corresponds to a droplet with a negative effective surface tension. The other solution for $\lambda = 0$ corresponds to

$$\tilde{\zeta} = \alpha_I \equiv -\frac{2(2l+1)\tilde{\eta}}{c_0}\left(\frac{D}{R_0} + \frac{D_b R_0 k_{\text{off}}}{(l+1)(D_b l + k_{\text{on}}R_0)}\right) \tag{S.21}$$

as given in the text. We see that the expression for $\lambda$ in terms of $l$ is complicated. However, since $\alpha_I$ does not depend on the bare surface tension $\gamma_0'$ the instability threshold does not depend on the deformation. Thus, we can find an approximate expression for $\lambda$ by solving $\dot{R} = 0$ for $\delta R_{lm}$, which gives:

$$\delta R_{lm}^* = \frac{\tilde{\zeta}R_0\delta c_{lm}}{2\gamma_0'(l+2)(l-1)}. \tag{S.22}$$

Substituting this for $\delta R_{lm}$ in equation (1) we find:

$$\frac{d\delta c_{lm}^*}{dt} = -\left[\frac{l(l+1)}{R_0^2}\left(\frac{\tilde{\zeta}c_0 R_0}{2(2l+1)\tilde{\eta}} + D\right) + \frac{D_b k_{\text{off}}l}{D_b l + k_{\text{on}}R_0}\right]\delta c_{lm}^* \equiv \lambda^*\delta c_{lm}^*. \tag{S.23}$$

The resulting equation for $\partial\lambda^*/\partial l = 0$ is a fifth order polynomial in $l$, this simplifies to a cubic equation in the limit $k_{\text{off}} \to 0$. The solution to this cubic equation is the analytical line plotted in figure 2. As shown in equation (??), the activity threshold does not increase linearly with $l$ if $k_{\text{off}} \neq 0$.

In 2-dimensions, the general solutions for flow around a circle are:

$$v_r = \begin{cases} \sum_{k=1}^{\infty} \left(a_k^{(0)} r^{k+1} + a_k^{(1)} r^{k-1}\right) e^{ik\theta} & \text{if } r < R_0 \\ \sum_{k=1}^{\infty} \left(b_k^{(0)} r^{-k+1} + b_k^{(1)} r^{-k-1}\right) e^{ik\theta} & \text{if } r > R_0 \end{cases} \quad (S.24)$$

$$v_\theta = \begin{cases} i\sum_{k=1}^{\infty} \left[(k+2)a_k^{(0)} r^{k+1} + k a_k^{(1)} r^{k-1}\right] e^{ik\theta}/k & \text{if } r < R_0 \\ -i\sum_{k=1}^{\infty} \left[(k-2)b_k^{(0)} r^{-k+1} + k b_k^{(1)} r^{-k-1}\right] e^{ik\theta}/k & \text{if } r < R_0 \end{cases} \quad (S.25)$$

$$P = \begin{cases} P_0 + 4\eta_0 \sum_{k=1}^{\infty} (k+1) a_k^{(0)} r^k e^{ik\theta}/k & \text{if } r < R_0 \\ P_0^{\text{ext}} + 4\eta_1 \sum_{k=1}^{\infty} (k-1) b_k^{(0)} r^{-k} e^{ik\theta}/k & \text{if } r > R_0, \end{cases} \quad (S.26)$$

where $r$ and $\theta$ are plane polar coordinates.

Applying the perturbation to the concentration and solving the boundary conditions of continuous velocity and force balance at the interface we can solve for the constants $a_k^{(i)}, b_k^{(i)}$ and find:

$$a_k^{(0)} = -\frac{k}{4R_0^{k+1}(\eta + \eta^{\text{ext}})} \left(\tilde{\zeta}\delta c_k - (k-1)\gamma_0' \frac{\delta R_k}{R_0}\right) \quad (S.27)$$

$$a_k^{(1)} = \frac{k}{4R_0^{k-1}(\eta + \eta^{\text{ext}})} \left(\tilde{\zeta}\delta c_k - (k+1)\gamma_0' \frac{\delta R_k}{R_0}\right) \quad (S.28)$$

$$b_k^{(0)} = -\frac{kR_0^{k-1}}{4(\eta + \eta^{\text{ext}})} \left(\tilde{\zeta}\delta c_k + (k+1)\gamma_0' \frac{\delta R_k}{R_0}\right) \quad (S.29)$$

$$b_k^{(1)} = \frac{kR_0^{k+1}}{4(\eta + \eta^{\text{ext}})} \left(\tilde{\zeta}\delta c_k + (k-1)\gamma_0' \frac{\delta R_k}{R_0}\right) \quad (S.30)$$

On substitution of these into the dynamic equations for $c$ and $R$, we see that the deformation does not depend on $\delta c_k$, and hence the droplet shape is always stable as a circle for positive effective surface tension ($\gamma_0' > 0$). This means that the only instability comes from the dynamics of $c$ on the surface:

$$\frac{\partial \delta c_k}{\partial t} = k\left\{-\frac{\tilde{\zeta}c_0}{2R_0\tilde{\eta}} - \frac{Dk}{R_0^2} - \frac{D_b k_{\text{off}}}{D_b k + k_{\text{on}} R_0}\right\} \delta c_k \equiv \lambda_{2D} \delta c_k \quad (S.31)$$

leading to the active threshold in equation (6). The maximally unstable mode can again be found by solving for the zeros in the gradient of the dispersion relation $\partial \lambda_{2D}/\partial k = 0$. The expressions for $k_{\max}$ are the solutions to the cubic equation:

$$\left[4DD_b^2\tilde{\eta}\right] k^3 + \left[D_b R_0 \left(\tilde{\zeta}D_b c_0 + 8Dk_{\text{on}}\tilde{\eta}\right)\right] k^2 + \left[2k_{\text{on}} R_0^2 \left(\tilde{\zeta}D_b c_0 + 2Dk_{\text{on}}\tilde{\eta}\right)\right] k + k_{\text{on}} R_0^3 \left(\tilde{\zeta}k_{\text{on}} c_0 + 2D_b k_{\text{off}}\tilde{\eta}\right) = 0 \quad (S.32)$$

When binding is omitted ($k_{\text{off}} = 0$) this reduces to a simple monotonic relation in the activity:

$$k_{\max,0} = -\tilde{\zeta}\frac{c_0 R_0}{4D\tilde{\eta}} \quad (S.33)$$

**Appendix B: Active Polar Droplet, perturbing the defect position**

One can predict the form of the $l = 1$ mode instability by displacing the defect position as outlined in the main text. The equilibrium polarisation field around such an arrangement is:

$$\boldsymbol{p}_{l=1} = \boldsymbol{e}_r + \delta z \frac{r - R_c}{r(R_0 - R_c)} \sin(\theta) \boldsymbol{e}_\theta. \quad (S.34)$$

This perturbation satisfies the equilibrium condition $\boldsymbol{h} = 0$ for the given boundary conditions which assume a radial hedgehog defect of radius $R_c$ centred at $r = 0$, and displaced a small distance $\delta z$ from the droplet centre. This polarisation field increases the overall passive free energy of the system, as it is energetically favourable for the defect

3

to be central. Using the expression for the bulk free energy in equation (5), and substituting in the expression for the polarisation (equation (??)) we get:

$$F_{\text{bulk}} = K \int_0^{2\pi} \int_0^{\pi} \int_{R_c}^{R_0 - \delta z \cos(\theta)} \sin(\theta) \mathrm{d}r \mathrm{d}\theta \mathrm{d}\phi \left\{ 1 - \delta z \frac{2(r - R_c) \cos(\theta)}{r(R_0 - R_c)} + \delta z^2 \left[ \frac{3r^2 - 6rR_c + 4R_c^2 + (r^2 - 2rR_c) \cos(2\theta)}{4r^2 (R_0 - R_c)^2} \right] \right\}$$

$$= F_0 + K \int_0^{2\pi} \int_0^{\pi} \sin(\theta) \mathrm{d}\theta \mathrm{d}\phi \left\{ -\delta z \cos(\theta) \left[ \frac{3(R_0 - R_c) + 2R_c \ln(R_c/R_0)}{R_0 - R_c} \right] + \delta z^2 \left[ \frac{7(R_0 - R_c) + 6R_c \ln(R_c/R_0)}{4(R_0 - R_c)^2} \right] \right.$$

$$\left. \delta z^2 \cos(2\theta) \left[ \frac{5R_0^2 - 9R_0 R_c + 4R_c^2 + 2R_0 R_c \ln(R_c/R_0)}{4R_0(R_0 - R_c)^2} \right] \right\} + O(\delta z^3)$$

$$\approx F_0 + \frac{4\pi K \delta z^2}{3R_0(R_0 - R_c)^2} \left[ (R_0 - R_c)(4R_0 + R_c) + 4R_0 R_c \ln(R_c/R_0) \right] \tag{S.35}$$

where $F_0 = 4K\pi(R_0 - R_c)$ is the bulk free energy for the unperturbed polarisation field. Thus, the net increase in free energy from this perturbation ($F_{\text{bulk}} - F_0$) can be written

$$\Delta F_{\text{bulk}} \approx \frac{4\pi K \delta z^2}{3R_0 (1 - \epsilon^2)} \left[ 4 - 3\epsilon - \epsilon^2 + 4\epsilon \ln(\epsilon) \right] \tag{S.36}$$

where $\epsilon = R_c/R_0$, as given in equation (8) of the text. This is positive for all $R_c < R_0$ and converges to a fixed value at small $\epsilon$.

Introducing a small activity, the active force for this polarisation field to leading order in $\delta z$:

$$\boldsymbol{f}^{\text{act}} = \nabla \cdot \underline{\underline{\sigma}}^{\text{act}} \approx 2\zeta \left[ -\frac{1}{r} + \delta z \frac{(r - R_c) \cos(\theta)}{r^2 (R_0 - R_{c0})} \right] \hat{\boldsymbol{r}} + \zeta \delta z \frac{(3r - 2R_c) \sin(\theta)}{r^2 (R_0 - R_{c0})} \hat{\boldsymbol{\theta}} \tag{S.37}$$

Solving the active part of the Stokes equation $\eta \nabla^2 \boldsymbol{v} + \boldsymbol{f}^{\text{act}} - \nabla P = 0$ for an incompressible flow we impose the boundary conditions of velocity and stress continuity with the external fluid:

$$\boldsymbol{v}^{\text{ext}}\big|_{r=R} - \boldsymbol{v}\big|_{r=R} = 0 \tag{S.38}$$

$$\hat{\boldsymbol{n}} \cdot \left[ \underline{\underline{\sigma}}^{\text{act}} + \sigma^{\text{visc}} - 2\eta_1 \underline{\underline{u}}^{\text{ext}} \right] \big|_{r=R} = \boldsymbol{F}_{lm} + (P - P^{\text{ext}}) \hat{\boldsymbol{n}} \big|_{r=R}. \tag{S.39}$$

These boundary conditions set the values of the constants for the general part of the solutions ($a_{10}^{(0)}$, $a_{10}^{(1)}$, $b_{10}^{(0)}$, $b_{10}^{(1)}$ from equations (??) and (??) for $l = 1, m = 0$). Thus, we arrive at an expression for the flow field:

$$\boldsymbol{v} \approx \frac{\zeta \delta z}{6R_0(R_0 - R_{c0})\eta_0} \left\{ \left[ R_0^2 \left( \frac{7\eta_0 + 3\eta_1}{3\eta_0 + 2\eta_1} \right) - 3rR_0 + 3r^2 \left( \frac{\eta_0 + \eta_1}{3\eta_0 + 2\eta_1} \right) \right] \cos(\theta) \hat{\boldsymbol{r}} \right. \tag{S.40}$$

$$\left. - \left[ R_0^2 \left( \frac{7\eta_0 + 3\eta_1}{3\eta_0 + 2\eta_1} \right) - \frac{9}{2} rR_0 + 6r^2 \left( \frac{\eta_0 + \eta_1}{3\eta_0 + 2\eta_1} \right) \right] \sin(\theta) \hat{\boldsymbol{\theta}} \right\} \tag{S.41}$$

The initial stability of the defect position will be determined by whether the defect is advected along its perturbation direction, relative to the velocity of the drop. The active part of this velocity is calculated by the integral over the defect $(1/V) \int_S (\boldsymbol{v}_{\text{act}} \cdot \hat{\boldsymbol{n}}) r r^2 \sin(\theta) \, \mathrm{d}\theta \mathrm{d}\phi$ at $r = R_c$, where $V$ is the defect volume enclosed by the surface $S$, this gives:

$$\boldsymbol{v}_c \approx \zeta \delta z \frac{\eta_0 (7 - 9\epsilon + 3\epsilon^2) + 3\eta_1 (1 - \epsilon)^2}{6\eta_0 (1 - \epsilon)(3\eta_0 + 2\eta_1)} \hat{\boldsymbol{z}} \tag{S.42}$$

where $\epsilon = R_{c0}/R_0$. The velocity of the interface is the result of the same integral over the surface $\Sigma$ at $r = R_0$:

$$\boldsymbol{v}_{\text{drop}} \approx \frac{\zeta \delta z}{6(1 - \epsilon)(3\eta_0 + 2\eta_1)} \hat{\boldsymbol{z}} \tag{S.43}$$

Finally, the relative velocity of the defect is given by $\boldsymbol{v}_c - \boldsymbol{v}_{\text{drop}}$:

$$\boldsymbol{v}_c - \boldsymbol{v}_{\text{drop}} \approx \zeta \delta z \frac{\eta_0 (2 - \epsilon) + \eta_1 (1 - \epsilon)}{2\eta_0 (3\eta_0 + 2\eta_1)} \hat{\boldsymbol{z}} \tag{S.44}$$

as in the main text in equation (8).

This result can be reproduced in 2-dimensions also. The form of the defect displacement is identical to the 3D case in plane polar coordinates $(r, \theta)$ (except the perturbation is along $x$ using this convention). Then the polarisation is:

$$\boldsymbol{p}_{2D} = \hat{\boldsymbol{e}}_r - \frac{\delta z(r^2 - R_c)}{r(R_0^2 - R_{c0}^2)} \sin(\theta) \hat{\boldsymbol{e}}_\theta. \tag{S.45}$$

The resulting defect speed and free energy change calculated as before are as follows:

$$\Delta v_{2D} = \frac{\zeta \delta x}{8\eta_0} \left[ \left( \frac{2\eta_0 + \eta_1}{\eta_0 + \eta_1} \right) - \frac{2\epsilon^2}{1 - \epsilon^2} \ln(\epsilon) \right], \tag{S.46}$$

$$\Delta F_{2D} = \frac{K\pi \delta x^2}{2R_0^2(1-\epsilon^2)^2} \left( 3 - 3\epsilon^2 + 2\epsilon^2 \ln(\epsilon) \right). \tag{S.47}$$

These predict similar behaviour as the 3D case, again with an extensile $\zeta > 0$ instability, and an increase in total bulk free energy.

**Appendix C: Strong Anchoring Approximation for Active Polar Drop. Calculation of flow field for $l > 1$.**

A general perturbation to the polarisation field that preserves $|p| = 1$ can be written in spherical coordinates as:

$$\boldsymbol{p} = \hat{\boldsymbol{r}} + \sum_{l=2}^{\infty} \sum_{m=-l}^{l} f_{lm}(r) \delta p_{lm}^{(1)}(t) \left( \frac{\partial Y_l^m}{\partial \theta} - \frac{1}{\sin(\theta)} \frac{\partial Y_l^m}{\partial \phi} \right) \hat{\boldsymbol{e}}_\theta + g_{lm}(r) \delta p_{lm}^{(2)}(t) \left( \frac{\partial Y_l^m}{\partial \theta} + \frac{1}{\sin(\theta)} \frac{\partial Y_l^m}{\partial \phi} \right) \hat{\boldsymbol{e}}_\phi.$$

In order to find a functional form for $f_{lm}(r)$ and $g_{lm}(r)$ one would need to calculate the resulting flows in terms of these functions, and solve the polarisation dynamics equations for these functions. We find that this is not an analytically tractable problem. Instead, we look for a suitable ansatz for the form of $f_{lm}(r)$ and $f_{lm}(r)$. First, we consider the limit of strong anchoring, this gives the boundary condition $\boldsymbol{p} = \hat{\boldsymbol{n}}$ at $r = R$, meaning (to linear order) $f_{lm}(R_0)\delta p_{lm}^{(1)}(t) = -\delta R_{lm}(t)/R_0$ and $g_{lm}(R_0)\delta p_{lm}^{(2)}(t) = 0$. Secondly, as discussed in the text, we consider the case where the polarisation in the bulk is strongly coupled to that at interface. An intuitive assumption may be to solve for $h = 0$ to first order, as we did for the $l = 1$ mode, but this yields a result with irrational powers of $r$. If we consider instead the polarisation field as rigid spokes, the deformation is described by $f_{lm}(r) = -\delta R_{lm}(t)/(R_0 \delta p_{lm})$ and $g_{lm}(r) = 0$, in other words there is no $r$-dependence. We find that this assumption leads to physically sensible results for a wide parameter range and it results in simple to understand results which reflect qualitatively what is observed in simulations. The important outcome of this is that, as long at the perturbation to the polarisation is small (and does not go to infinity at the origin), regardless of our choice of the $r$ dependence of the perturbation, we see that the active part of the flow always deforms the interface in the same way. In other words contractile activity acts to promote the deformation, while extensile acts to reduce the deformation.

Using this perturbation (equation (7)) we can first calculate partial solutions for the velocity and pressure that balance the force terms in the Stokes equation ($\eta_0 \nabla^2 \boldsymbol{v} + \nabla \cdot \sigma_{\text{act}} + \sigma_{\text{dis}}$), the partial solution for the velocity is:

$$\boldsymbol{v}_{\text{part}} = -\frac{\delta R_{lm}}{R_0 \eta_0} \left\{ \zeta \frac{l(l+1)r}{(l+3)(l-2)} + \frac{K}{2r} \left( (1+\nu)l(l+1) + 2(1-\nu) \right) \right\} Y_l^m(\theta, \phi) \hat{\boldsymbol{e}}_r$$
$$-\frac{\delta R_{lm}}{R_0 \eta_0} \left\{ \zeta \frac{3r}{(l+3)(l-2)} + \frac{K}{2l(l+1)r} \left( (1+\nu)l(l+1) + 2(1-\nu) \right) \right\} r \nabla Y_l^m(\theta, \phi) \tag{S.48}$$

Note that the active part of this flow solution appears to diverge for $l = 2$, however if we specifically solve for $l = 2$ we see that this is because the actual form of this is proportional to $r^2 \log(r)$ (the solution containing the term of this form is plotted in figure 4). However, using the general form above results in the same eventual result for the stability analysis. These partial solutions can then be added to the general solutions for flow around a sphere (equations (**??**), (**??**) and (**??**)) so that one can solve the flow and stress continuity boundary conditions at $r = R$:

$$\boldsymbol{v}^{\text{ext}} \big|_{r=R} - \boldsymbol{v} \big|_{r=R} = 0 \tag{S.49}$$

$$\hat{\boldsymbol{n}} \cdot \left[ \underline{\underline{\sigma}}_{\text{tot}} - 2\eta_1 \underline{\underline{u}}^{\text{ext}} \right] \big|_{r=R} = \boldsymbol{F}_{lm} + (P - P^{\text{ext}}) \hat{\boldsymbol{n}} \big|_{r=R}. \tag{S.50}$$

Note that the internal stress at the interface is no longer just the viscous stress, it also contains $\sigma_{\text{act}}$ and $\sigma_{\text{dis}}$. Also, we assume a constant surface tension $\gamma_0$, so $\boldsymbol{F}_{lm}$ now is as defined in the text for the active interface but with $c = 0$.

5

Solving the above boundary conditions for $a_{lm}^{(i)}$ and $b_{lm}^{(i)}$ we can substitute the full velocity solution into the equation for $\dot{R}$ and find the resulting stability parameter. Note that because we are assuming a polarisation $\boldsymbol{p}$ that is fixed by the anchoring, it is the radius dynamics that can be used to determine the stability. The resulting stability parameter $\lambda$ is:

$$\begin{aligned}
\lambda = &-\frac{K}{2R_0^2 \eta_{\text{eff}}^2} \{l(l-1)(l+2)\nu\left[3\eta_1 + 8l(l+1)(l+2)\tilde{\eta}\right] \\
&+ l\left[2\eta_0(l+1)(2l(l(l+5)+7)-5) + \eta_1(l(l(4l(l(l+6)+12)+37)+5)+2)\right]\} \\
&- \zeta \left\{ \frac{l(l+1)(\eta_0(l-1)(l(4l+9)+3) + \eta_1(l(l(4l+5)-3)-3))}{(l+3)\left(\eta_1 + 2\eta_0\left(l^2-1\right) + 2\eta_1 l^2\right)\left(\eta_0(2l(l+2)+3) + 2\eta_1 l(l+2)\right)} \right\} \\
&- \gamma_0 \left\{ \frac{(l-1)l(l+1)(l+2)(2l+1)(\eta_0+\eta_1)}{R_0 \left(\eta_1 + 2\eta_0\left(l^2-1\right) + 2\eta_1 l^2\right)\left(\eta_0(2l(l+2)+3) + 2\eta_1 l(l+2)\right)} \right\}
\end{aligned} \quad (S.51)$$

where $\eta_{\text{eff}}^2 = \left(\eta_1 + 2\eta_0\left(l^2 - 1\right) + 2\eta_1 l^2\right)\left(\eta_0(2l(l+2)+3) + 2\eta_1 l(l+2)\right)$. Then, the resulting contractile active threshold is found by solving $\lambda = 0$ for $\zeta$. The resulting dispersion relation is very complex in terms of $l$, and hence we are able to do little analytically to quantify it. However, we find that for a large parameter range the active threshold tends to increase in magnitude with mode number $l$, as one would expect. However, the distortion contribution (proportional to $K$) to the flow field in some instances ($\nu < -1$) appears to destabilise the drop at very large $l$. This is likely due to the assumption we make about the form of the polarisation perturbation, however the results appear to hold well for the $l = 2$ mode, which we observe in the simulations as a symmetric deformation of the droplet.

In order to find this in 2D we work through exactly the same set of operations in polar coordinates, except that we assume that the polarisation scales as $p_\theta \propto r$, as using $r^0$ results in infinite passive contributions. The resulting activity threshold is plotted in figure 5 for the simulation parameters using the $k = 2$ mode.



\end{document}